\documentclass[%
 reprint,
 amsmath,amssymb,
 aps,
 prd,
]{revtex4-1}

\usepackage{fitpar}
\usepackage{fiterr}
\usepackage{variables}
\usepackage[colorlinks = true,
            linkcolor = cyan,
            urlcolor  = blue,
            citecolor = red,
            anchorcolor = blue]{hyperref}\usepackage{color}

\usepackage{graphicx}
\usepackage{dcolumn}
\usepackage{bm}
\usepackage{url}
\usepackage{natbib}
\usepackage{siunitx}
\usepackage{lipsum}
\usepackage{amsmath}
\usepackage{amssymb}
\usepackage{titlesec}
\titlespacing\section{0pt}{6pt plus 4pt minus 2pt}{4pt plus 2pt minus 2pt}
\titlespacing\subsection{0pt}{6pt plus 4pt minus 2pt}{4pt plus 2pt minus 2pt}
\titlespacing\subsubsection{0pt}{6pt plus 4pt minus 2pt}{4pt plus 2pt minus 2pt}
\usepackage{physics}

\usepackage{booktabs}
\usepackage[normalem]{ulem}
\useunder{\uline}{\ul}{}
\usepackage{multirow}
\usepackage{hhline}

\usepackage[margin=1in]{geometry}
\usepackage{array}
\newcolumntype{C}{>{$}c<{$}}
\AtBeginDocument{
\heavyrulewidth=.08em
\lightrulewidth=.05em
\cmidrulewidth=.03em
\belowrulesep=.65ex
\belowbottomsep=0pt
\aboverulesep=.4ex
\abovetopsep=0pt
\cmidrulesep=\doublerulesep
\cmidrulekern=.5em
\defaultaddspace=.5em
}

\usepackage[mathlines]{lineno}

\newcommand{\poten}{$^{210}$Po}

\newcommand{\pofour}{$^{214}$Po}
\newcommand{\pbsix}{$^{206}$Pb}
\newcommand{\bifour}{$^{214}$Bi}
\newcommand{\rntwo}{$^{222}$Rn}
\newcommand{\rasix}{$^{226}$Ra}
\newcommand{\pbten}{$^{210}$Pb}

\newcommand{\sry}{$^{90}$Sr/$^{90}$Y}

\newcommand{\lntwo}{LN$_2$}
\newcommand{\ntwo}{N$_2$}

\newcommand*\Ei{\mathrm{Ei}}

\newcommand{\surfaceactivity}{100\,\SI{}{\becquerel/\centi\meter\squared}}
\newcommand{\us}{\,\SI{}{\micro\second}}
\newcommand{\um}{\,\SI{}{\micro\meter}}
\newcommand{\ug}{\,\SI{}{\micro\gram}}

\usepackage{xcolor}

\begin{document}

\title{Surface background suppression in liquid argon dark matter detectors using a newly discovered time component of tetraphenyl-butadiene scintillation}

\author{Chris Stanford}
\email{cstan4d@stanford.edu. Current address: Physics Department, Stanford University, Stanford, California 94305, USA}
 \affiliation{Department of Physics, Princeton University, Princeton, New Jersey 08544, USA} %

\author{Shawn Westerdale}%
\altaffiliation{Current address: Department of Physics, Carleton University, Ottawa, Ontario, Canada}
 \affiliation{Department of Physics, Princeton University, Princeton, New Jersey 08544, USA} 
\author{Jingke Xu}%
 \altaffiliation{Current address: Lawrence Livermore National Laboratory, Livermore, California 94550, USA}
 \affiliation{Department of Physics, Princeton University, Princeton, New Jersey 08544, USA} 
\author{Frank Calaprice}%
 \affiliation{Department of Physics, Princeton University, Princeton, New Jersey 08544, USA} 

\newcommand{\test}{}

\date{\today}

\begin{abstract}
Decays of radioisotopes on inner detector surfaces can pose a major background concern for the direct detection of dark matter. 
While these backgrounds are conventionally mitigated with position cuts, these cuts reduce the exposure of the detector by decreasing the sensitive mass, and uncertainty in position determination may make it impossible to adequately remove such events in certain detectors.
In this paper, we provide a new technique for substantially reducing these surface backgrounds in liquid argon (LAr) detectors, independent of position cuts. 
These detectors typically use a coating of tetraphenyl-butadiene (TPB) on the inner surfaces as a wavelength shifter to convert vacuum ultraviolet (VUV) LAr scintillation light to the visible spectrum.
We find that TPB scintillation contains a component with a previously unreported exceptionally long lifetime ($\sim$ms). We discovered that this component differs significantly in magnitude between alpha, beta, and VUV excitation, which enables the use of pulse shape discrimination to suppress surface backgrounds by more than a factor of $10^3$ with negligible loss of dark matter sensitivity. We also discuss how this technique can be extended beyond just LAr experiments.

\end{abstract}

\maketitle

Direct detection dark matter experiments look for signals produced by dark matter particles, such as Weakly Interacting Massive Particles (WIMPs) that may scatter off target atoms in a detector and produce nuclear recoils (NRs). These interactions are expected to be rare, so  WIMP detectors are required to have very low backgrounds. 

One source of background can come from radioactive decays on the inner surfaces of the detector. When these isotopes decay, their decay products may produce a NR signal in the target volume. Of particular concern is \poten, a descendant of long-lived \pbten\ in the \rntwo\ decay chain. Since \rntwo\ is naturally present in the air, \poten\ can appear on detector surfaces many years after their exposure to air during construction or handling.

Absolutely clean surfaces are unattainable, so surface backgrounds are commonly found in dark matter experiments \cite{ref:CDMSIIFinal,ref:LUXFirst,ref:XENON1TFirst,ref:DEAPfirst,ref:DS532day}. These backgrounds are ordinarily suppressed by making position-based cuts.
Doing so involves rejecting all the events that occur near the surfaces, effectively reducing the volume of the target to an inner ``fiducial" volume, which reduces the detector's total exposure. 
Depending on the position resolution of the dark matter detector, substantial fiducial cuts may be needed. Furthermore, experiments that have regions in their detector with systematically poor position reconstruction may find it impractical to fiducialize enough to fully suppress these backgrounds.

Liquid argon (LAr) detectors are actively used for the direct detection of both light~\cite{ref:DS50LowMass,ref:DS50SubGeV} and heavy~\cite{ref:DEAPfirst,ref:DS532day} WIMP dark matter. LAr features a high scintillation light yield and a scintillation time profile that differs between nuclear and electronic recoils, which allows pulse shape discrimination (PSD) to be used to separate and suppress backgrounds caused by gammas or betas scattering off atomic electrons.

Upon excitation or ionization, LAr emits scintillation photons in the vacuum ultraviolet (128\,nm), which are difficult to detect directly. 
For more efficient light collection, LAr experiments usually apply a thin layer of wavelength shifter (WLS) to the inner surfaces to absorb the scintillation photons and re-emit them at longer wavelengths. 
Unfortunately, the re-emission by the WLS is approximately isotropic and can obscure the original location of the event within the detector. This results in lower position resolution, necessitating larger fiducial cuts to avoid surface backgrounds. For example, the DEAP-3600 LAr dark matter experiment loses one third of its 3.3\,tonne target mass to fiducial cuts~\cite{ref:DEAPfirst}.

A common wavelength shifter for LAr detectors is 1,1,4,4-tetraphenyl-1,3-butadiene (TPB). 
TPB is also a scintillator~\cite{ref:TPBAlphaPollman}, and has been shown to produce light under the direct excitation of alpha and beta particles~\cite{ref:TPBSegreto}. 
A radioisotope that decays on the surface may therefore produce a signal in the TPB, in the LAr, or in both.
This scintillation signal may be used to help identify surface background events, as to be discussed in Section~\ref{sec:decaymodes}. 

\begin{figure*}[th!]
\includegraphics[width=0.8\textwidth]{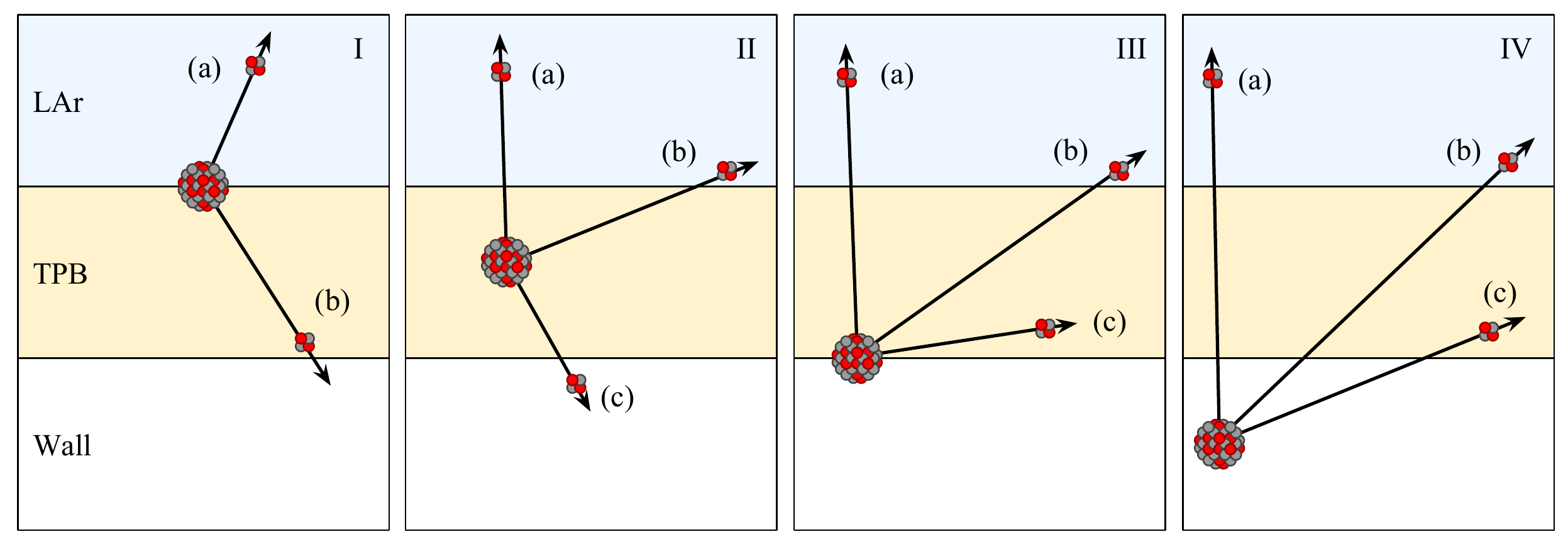}
\caption{Possible alpha tracks, or decay modes, on the surface of a liquid argon detector depend on the location of the decaying radioisotope. Of particular concern is decay mode I(b), which has a low energy signal produced in the liquid by the recoiling nucleus, as well as decay modes (II-IV)(b), which have low energy signals produced in the liquid by a degraded alpha. Decay modes (II-IV)(c) produce scintillation only in the TPB, but this can still be a concern for experiments that do not rely on charge collection.}
\label{fig:decay_kin}
\end{figure*}

In this paper, we measure the scintillation pulse shape of a typical surface alpha decay in a LAr detector. 
We also measure the pulse shape of the TPB response to alphas, betas, and 128\,nm photons.
We measure these pulse shapes out to the millisecond timescale, significantly longer than previous experiments.
We find that the TPB response contains a previously unreported milliseconds-long component with a magnitude that depends strongly upon the type of excitation.
Using this discovery, we demonstrate that PSD can be used to suppress surface backgrounds in LAr dark matter detectors by more than a factor of $10^3$ with negligible loss of nuclear recoil acceptance or detector sensitivity.

\section{SURFACE DECAY MODES\label{sec:decaymodes}}

As we reported in Ref.~\cite{ref:PbRecoil}, alpha decays on detector surfaces can produce NR signals. Due to the tendency of \poten\ to appear on inner detector surfaces for years after their exposure to air, it is the most relevant alpha-emitting isotope for WIMP search experiments. The decay of \poten\ produces a 5.3\,MeV alpha particle and a 103\,keV recoiling \pbsix\ nucleus.

When this decay occurs on a LAr detector surface coated with TPB, a variety of signals can be produced depending on the location of the isotope and the emission direction of the decay products. The possible modes for the decay are illustrated in Fig.~\ref{fig:decay_kin}. We examine each mode and discuss its potential for creating a background of concern for a dark matter search:

\vspace{0.2cm}
\noindent\textbf{(I-IV)(a)}: The alpha deposits a large fraction of its energy ($\sim$MeV) in the LAr. This mode does not pose a background for detectors looking for the sub-100\,keV nuclear recoils expected from WIMPs.

\noindent\textbf{I(b)}: The alpha is directed toward the wall, and it deposits some or all of its energy in the TPB, depending on the angle. The recoiling nucleus can be ejected from the surface into the LAr and produce a low energy nuclear recoil, a potential background. This recoil signal has been characterized in Ref.~\cite{ref:PbRecoil}.

\noindent\textbf{(II-IV)(b)}: The alpha's energy is degraded as it passes through the surface material and has the potential to be degraded enough to fall in the detector's WIMP search region.

\noindent\textbf{(II-IV)(c)}: No signal is produced in the LAr, but the alpha will still cause scintillation in the TPB.
The size of this signal has been measured to be in the dark matter search region of interest~\cite{ref:PbRecoil}. It could pose a background for single-phase experiments that do not depend on charge collection from interactions in the LAr.

This examination of the surface decay kinematics reveals that each mode carrying a risk of background features some amount of TPB scintillation from the alpha. The unique response of TPB to alpha particles (that we report below) therefore provides a means of suppressing all of these modes.

\section{APPARATUS}

\begin{figure}[tp]
\includegraphics[width=0.4\textwidth]{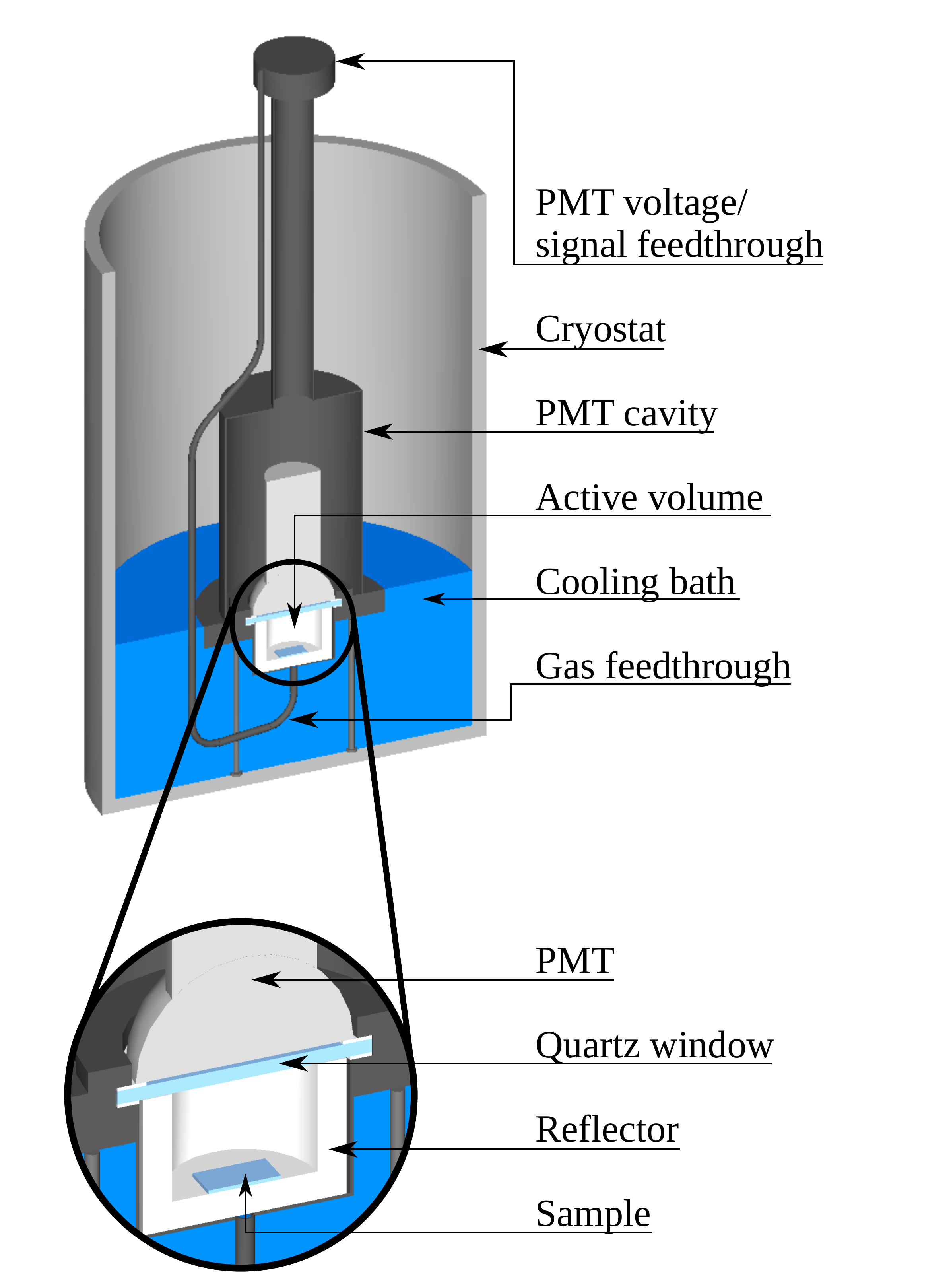}
\caption{A cutaway view of the detector. The zoomed inset shows the lower chamber containing the sensitive volume where the samples are placed.}
\label{fig:detector_markup}
\end{figure}

We built the detector, depicted in Fig.~\ref{fig:detector_markup}, to perform scintillation studies of TPB under different radioactive excitations at both room and cryogenic temperatures. It consists of two stainless steel chambers separated by a quartz window. The lower chamber, a cylinder with a diameter of 7\,cm and height of 5\,cm, is lined with 1\,cm-thick high-reflectivity PTFE, forming a reflecting cup. Samples containing TPB are placed into this cup with radioactive sources, and the scintillation light is collected by a single Hamamatsu R11065 photomultiplier tube (PMT) hosted in the upper chamber, directed downward at the quartz window. 
The small size of the reflecting cup was chosen to reduce the number of reflections of the scintillation light and thus enhance the overall light collection efficiency. 
The quartz window makes a compression seal to the lower chamber with a PTFE gasket; this feature enabled us to make scintillation measurements with the sample in vacuum while operating the PMT in a gas environment. 

Both chambers are connected to a Pfeiffer turbomolecular vacuum pump~\cite{ref:turbopump}, which is capable of producing a $10^{-6}$\,mbar vacuum in the lower chamber. The system is also connected to a high purity (99.999\%) argon gas bottle through a SAES getter~\cite{ref:SAES}, allowing the chambers to be filled with scintillation-grade argon. 
The detector is mounted on three legs, so it can stand upright, and can be lowered into the bottom of a double-walled dewar. The dewar can then be filled with liquid nitrogen (\lntwo) or LAr to perform cryogenic measurements.
The dewar is surrounded by lead bricks to reduce the external gamma backgrounds.

The data acquisition system consists of an analog amplifier ($\times$10) and a CAEN V1720 analog-to-digital converter (ADC) with a 4\,ns sampling rate~\cite{ref:v1720}. The digital signal is read out by a PC via a CAEN A2818 optical controller and recorded with the Daqman software~\cite{ref:daqman}. Daqman is also used to process the raw waveforms and analyze the data.

More details of the apparatus can be found in~\cite{ref:thesis_chris_stanford_radose}.

\section{MEASUREMENTS\label{sec:measurements}}

\begin{figure}[tp]
\includegraphics[width=0.5\textwidth]{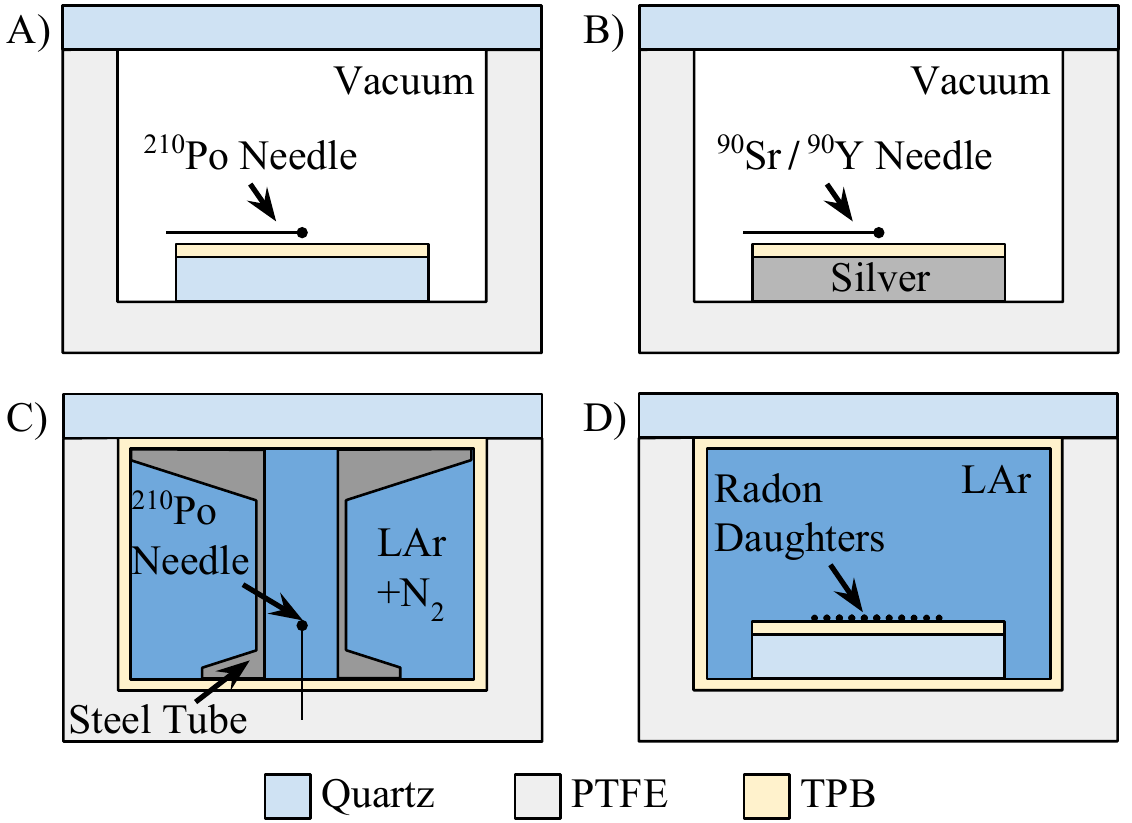}
\caption{Cross sections of the lower chamber (Fig.~\ref{fig:detector_markup} inset) for the different TPB response measurements. A) Alphas from a \poten\ source. B) Betas from a \sry\ source. C) 128\,nm photons from LAr scintillation. D) Surface alpha decays in LAr.}
\label{fig:schematics}
\end{figure}

We carried out a comprehensive set of measurements to characterize the TPB response under different excitation conditions. Correspondingly, the lower detector chamber was configured in different ways, as illustrated in Fig.~\ref{fig:schematics}(A-D). The scintillation response of TPB was measured for alphas and betas in a vacuum (A\&B) and for 128\,nm photons in LAr (C). Finally, we measured the full signal expected from a surface alpha decay in a LAr detector (D) by depositing radon progeny, including the alpha-emitting isotope \pofour, on a TPB-coated slide and placing the slide in an active LAr volume.

\subsection*{Measurement A: Alphas (Fig.~\ref{fig:schematics}A)}
For the measurement of alpha-induced scintillation in TPB, a \poten\ alpha needle source was secured with PTFE tape to the top of a quartz slide coated with $205\pm10$\,\ug/cm$^2$ TPB. The surface density was computed by recording the weight of the slide with a precision scale before and after the deposition. The slide with the alpha source was placed in the detector, and the chamber containing the source was pumped down to a vacuum of approximately $10^{-3}$\,mbar to minimize possible scintillation contamination from nitrogen or other gases in the environment. 

To prepare the slide, it was placed in an ultrasonic bath with detergent for thirty minutes. Then it was rinsed, first with water then with isopropanol. Then the slide was left to bake in a vacuum oven at 120$^{\circ}$C for a minimum of five hours. After baking, the slide was placed in a vacuum chamber where the TPB coating was applied via evaporation. The same procedure was followed for all substrate materials used in the following studies.

The measurement was performed at two temperatures: ambient lab temperature (295~K) and \lntwo\ temperature (77~K). To obtain the \lntwo\ temperature data, the bottom half of the detector (the entirety of the lower chamber) was submerged in \lntwo. Measurements were recorded after the detector had been submerged for 1 hour to allow enough time for the TPB to reach equilibrium with the \lntwo. 

\subsection*{Measurement B: Betas (Fig.~\ref{fig:schematics}B)}

The TPB scintillation response to beta excitation was measured using a \sry\ needle source~\cite{ref:NeedleSources}, a nearly-pure beta emitter ($I_\gamma\sim10^{-4}$~\cite{ref:y90}) with Q-values of 546\,keV for $^{90}$Sr and 2279\,keV for $^{90}$Y.
The source was fastened with PTFE tape to a pure silver slide coated with TPB. Since betas have a smaller energy loss in TPB per unit distance than alphas, a thicker layer of TPB ($1290\pm50$\,\ug/cm$^2$) was used to capture more of the beta energy. The silver slide was chosen instead of quartz to prevent the release of Cerenkov light produced by the beta particles in the substrate.

The beta response was observed at lab temperature and \lntwo\ temperature in the same manner as the alpha measurements.

\subsection*{Measurement C: 128\,nm Photons (Fig.~\ref{fig:schematics}C)}

In this measurement, we evaluated the response of TPB to 128\,nm LAr scintillation light. 
Because LAr scintillation comprises a prompt component (lifetime of nanoseconds) and a slow component (lifetime of microseconds) that may obscure the time response of the TPB, we doped LAr with \ntwo\ so that it contained approximately 0.1\% \ntwo\ in order to quench the slow component.
According to the dedicated study of this phenomenon reported in~\cite{ref:N2doping}, the slow component disappears at the 500\,ppm level. 
The result of this doping is that the LAr scintillation pulses in this measurement would be fully prompt with no measurable slow component. Thus, any longer time structure in the measured pulses could be attributed to the response of the TPB.

As illustrated in Fig.~\ref{fig:schematics}C, we used a \poten\ source to excite LAr scintillation. The \poten\ needle source was stuck into the bottom of the PTFE cup, which was now coated with $265\pm50$\,\ug/cm$^2$ of TPB. The needle stood upright so that the alpha emission site (eye of the needle) was in the center of the volume, far from the TPB-coated surfaces (we calculated the range of 5.3 MeV alphas to be 45\,\um\ in LAr~\cite{ref:ASTAR}).
Note that the \poten\ alphas, when depositing their full energy in LAr as they do here, would produce enough scintillation light to saturate the electronics. 
To address this problem, we placed a hollow stainless steel tube with an inner diameter of 1.3\,cm around the needle source. The tube drastically reduced the light collection, as the 128\,nm LAr light had only a small solid angle through which to escape the tube and reach the window.

The procedure for filling the detector with LAr began by pumping the chambers down to a vacuum of $10^{-6}$\,mbar. Then the detector was placed into an empty dewar, which was then filled with LAr until the lower chamber was submerged. Subsequently, gaseous argon (GAr) was allowed to flow into the detector where it condensed into a liquid. This continued for 2 hours until the gas flow ceased, indicating that the lower chamber was fully filled with LAr. This procedure was later verified by visually observing the liquid level through the quartz window with the PMT and the upper chamber removed during cooling and filling.

\subsection*{Measurement D: Surface Decay in LAr (Fig.~\ref{fig:schematics}D)}

We directly measured the scintillation signals from surface alpha decay events in the LAr detector using the configuration illustrated in Fig.~\ref{fig:schematics}D. 
The alpha decays were produced by radon progeny deposited on the surface of a TPB coating on a quartz slide. 
The quartz slide was coated with $265\pm50$\,\ug/cm$^2$ of TPB (typical of a dark matter experiment~\cite{ref:DS50day}), and was left for ten hours in a chamber filled with argon and \rntwo\ gas obtained from a \rasix\ source. 
The initial activity of the radon progeny on the slides, right after its removal from the radon chamber, was measured with a Geiger counter to be about 500\,Bq, or \surfaceactivity.
The slide was then immediately placed in the LAr detector where the TPB-coated quartz window and PTFE cup remained in the same condition as used in Measurement C.
The detector was then closed, pumped and purged with GAr several times, cooled down, and filled with GAr using the same procedure described in Measurement C. This GAr was not doped with \ntwo, but rather passed through a SAES gas purifier~\cite{ref:SAES}. 

The alpha decay studied in this measurement was that of \pofour. While the \pofour\ decay produces an alpha with a higher energy than that of \poten, \pofour's short half-life of 164\us\ allows it to be tagged by its delayed coincidence with the preceding beta decay of \bifour\ for a substantial reduction in background. More details on this method can be found in \cite{ref:PbRecoil} and \cite{ref:thesis_chris_stanford}. The systematic effect introduced by our use of this more energetic decay is discussed in Section~\ref{sec:suppress}.

\section{ANALYSIS AND RESULTS\label{sec:results}}

\begin{figure}[tp]
\includegraphics[width=0.48\textwidth]{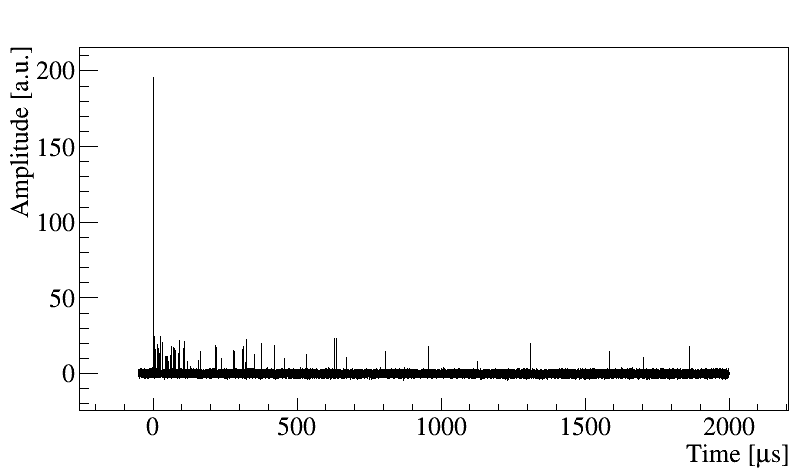}
\caption[A single waveform from alpha scintillation in TPB at \lntwo\ temperature]{A single waveform from alpha scintillation in TPB at \lntwo\ temperature. The prompt component occurs at Time=0 and has a lifetime of a few nanoseconds, while the delayed component has a lifetime of milliseconds. The small pulses throughout the delayed component correspond to the detection of single photoelectrons.}
\label{fig:single_waveform}
\end{figure}

In analyzing the TPB scintillation from \poten\ alphas (Measurement A), we discovered that a significant fraction of the scintillation light had a lifetime on the order of milliseconds. A sample waveform can be seen in Fig.~\ref{fig:single_waveform}, where the scintillation can be seen extending to the end of the 2\,ms acquisition window. This is orders of magnitude longer than the TPB scintillation lifetimes recorded in previous studies. These studies used shorter acquisition windows (10\us~\cite{ref:TPBSegreto} and 200\us~\cite{ref:TPBTemp}), which likely led to an underestimation of this component. While we find consistency in the time profile with previous measurements over the shorter timescales that they employed, our use of a 2\,ms-long window revealed that this component of TPB scintillation lasts much longer than previously understood. This long-lived component is consistent with concurrent results reported by Asaadi et al., who saw similar lifetimes in TPB dissolved in LAr and excited by a UV LED~\cite{ref:Asaadi}. 

To thoroughly characterize this milliseconds-long component, the individual waveforms for each configuration were summed to create average waveforms (AWs). Data selection cuts were applied to remove waveforms that contained additional pulses caused by signal pileup or PMT afterpulsing. These were defined as pulses outside of the trigger region $([-0.1,0.2]\us)$ crossing a discriminator threshold of 2.5 single photoelectrons.
Waveforms with greater than 95\% of their light arriving within 90\,ns of the trigger were rejected as Cherenkov backgrounds. Waveforms with less than 5\% of their light arriving within 90\,ns of the trigger were rejected as triggers on the tail of a previous signal. For Measurement D, the trigger occurred on the \bifour\ decay, with the pulse from the \pofour\ decay appearing later in the same waveform. In this case, waveforms were rejected if they contained a third pulse, and were also rejected if the second pulse occurred within 25\,us of the first in order to avoid contaminating the alpha signal with the tail of the beta signal. The systematic uncertainties introduced by varying these cut definitions were studied and found to have a negligible impact on the final results. 

\begin{figure}[tp]
\includegraphics[width=0.5\textwidth]{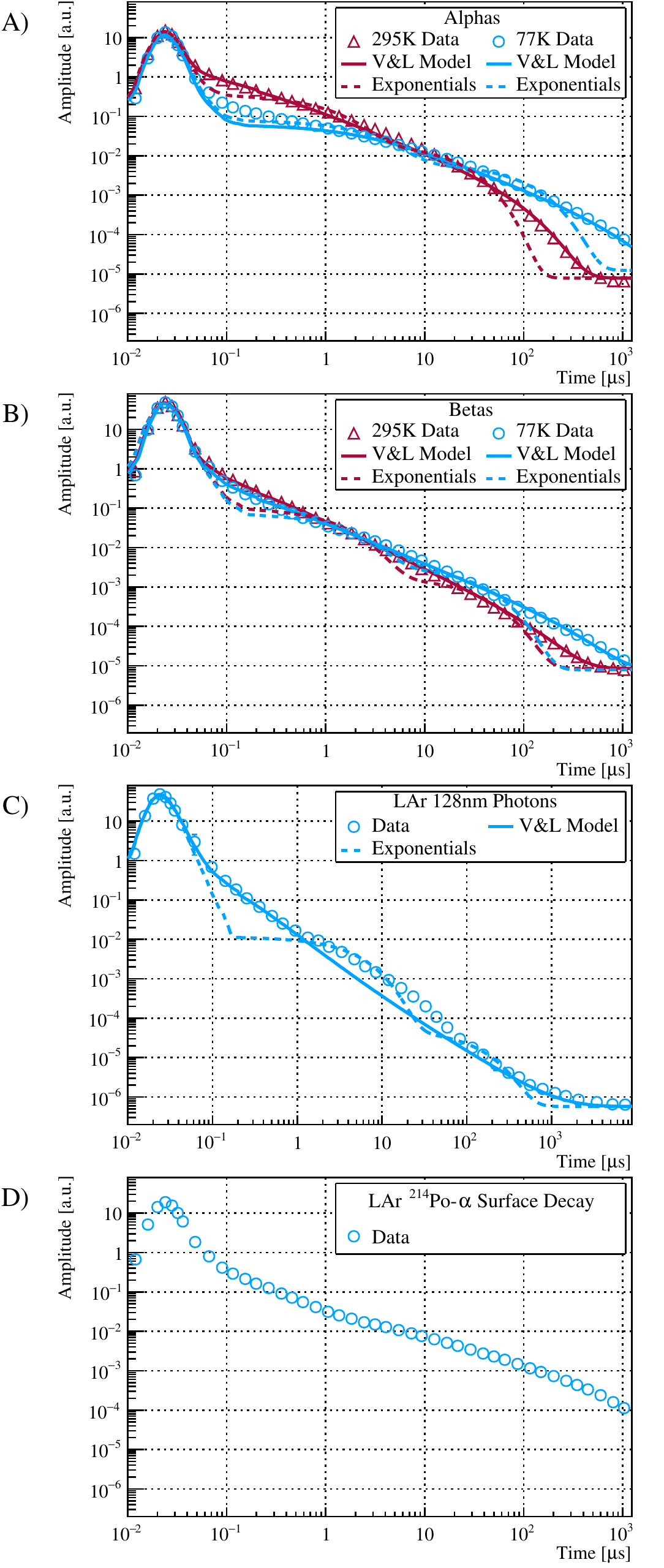}
\caption{The average waveforms of the TPB response to alphas (A), betas (B), 128\,nm photons (C), and surface alpha decays (D). In each case, following a prompt component with a lifetime of nanoseconds, evidence of a longer time component with a lifetime of milliseconds can be seen, and was found to be temperature dependent. A model (solid lines) developed by Voltz and Laustriat~\cite{ref:LongTailTheory} is used to fit the data (see Appendix~\ref{sec:model}). This model offers a better fit than a function of three exponentials (dashed lines) with the same number of free parameters.}
\label{fig:averages}
\end{figure}

The AWs exhibit both high intensity at short timescales and low intensity at long timescales, so for clarity they are displayed on log-log axes in Fig.~\ref{fig:averages}. The time axes of the AWs were re-binned into exponentially increasing bin sizes, such that the bin widths would appear uniform on a log axis. The AWs were then normalized to an integral of 1, taking into account the varying bin widths.

As illustrated in Fig.~\ref{fig:averages}, the AW for each measurement exhibits a clear milliseconds-long component (hereafter called the ``delayed" component) following the prompt component.
Several measures were undertaken to ensure that the delayed component was not caused by a systematic effect unrelated to the TPB. First, multiple PMTs were used to verify that it was not a result of a faulty light detector. Second, background runs were taken without a radiation source to quantify the baseline noise and dark count rate of the PMT, as well as external backgrounds in the LAr like gammas and cosmogenic muons.  Third, multiple substrates were tested under alpha excitation, including quartz, PTFE, and silver, to verify that the scintillation was coming only from the TPB and not from the substrate. Fourth, for every trigger of the data acquisition system, we recorded a minimum of 50\,\us\ prior to the trigger to ensure a clean baseline before an event.
With these tests carried out, we concluded that the delayed component remained unchanged with different PMTs, and no measurable scintillation was found from alpha excitation in the tested substrates. The constant baseline noise and dark rate were confirmed to occur at a level well below the delayed component.

Background contamination in these measurements was determined to be insignificant. 
In Measurements A\&B, TPB scintillation from external radioactivity was negligible due to the extremely small mass of the TPB coating.
The relevant background rate in the LAr configurations (measurements C\&D) was measured to be less than 1\,Hz, thanks to the lead shielding surrounding the detector and the unique signal signatures. Measurement C, which looks for MeV-scale signals from alpha depositions in LAr, eliminates this remaining external background by setting a high energy threshold. Measurement D eliminates this background with a \bifour--\pofour\ delayed coincidence search.

Fig.~\ref{fig:averages} also shows function fits to the AWs. Initial attempts to use a set of three exponentials (dotted lines) to fit the AWs failed. We were able to obtain better fits with the same number of free parameters by following the model (solid lines) proposed by Voltz and Laustriat~\cite{ref:LongTailTheory}. In this model, the TPB scintillation is described as two independent components: a prompt component that undergoes exponential decay with a lifetime on the order of nanoseconds, and a delayed component that undergoes an approximately power-law decay with an exponential cutoff on the order of milliseconds. Due to the complexity of the model, the detailed discussions are left in Appendix~\ref{sec:model}. 

Although the delayed component was observed in all measurements, the amplitude of this component was found to depend heavily upon the type of excitation. Using the scintillation model, the fraction of light in the delayed component for alphas, betas, and 128\,nm photons was evaluated to be approximately 80\%, 30\%, and 10\%, respectively. This difference enables strong discrimination between different interaction types and allows for powerful surface background suppression, as to be elaborated in the next section. 

The delayed component of the TPB scintillation also appears to depend on the temperature, as can be seen in Fig.~\ref{fig:averages} A\&B. The fitting of the scintillation model revealed that while the lifetime can be seen to increase at cryogenic temperature, the ratio of the prompt to the delayed component remains unchanged. This observation has implications for LAr detector simulations and previous measurements of the TPB light yield, as elaborated in Appendix~\ref{sec:TPBLY}.

\section{SURFACE BACKGROUND SUPPRESSION\label{sec:suppress}}

Due to the difference in TPB response between alphas and 128\,nm photons as discussed above, surface alpha background events produce pulses with a substantially larger fraction of light in the delayed component than NR events in the bulk LAr.
We can take advantage of this pulse shape difference to suppress surface alpha backgrounds in LAr dark matter detectors and maintain a much higher signal acceptance compared to the use of position cuts. 

To quantify this background rejection power, we compared the surface background events obtained in Measurement D to data published by DarkSide-50, a dark matter experiment with a LAr target mass~\cite{ref:DS70day}. DarkSide-50 uses a pulse shape parameter called $f_{90}$, defined as the integral of the first 90\,ns of the pulse divided by the integral of the entire pulse (for which DarkSide-50 uses a 7\us-long window), to distinguish NR events from the electron recoil events. The energy is measured in terms of the total number of photoelectrons (NPE) in the pulse collected by the PMTs. 
The NR signal acceptance contours in the $f_{90}$-NPE parameter space were determined using an in-situ neutron calibration in DarkSide-50~\cite{ref:DSCALIS}.

In this discussion, we will recalculate the NR acceptance region with an integration window ($W$) larger than 7\us\ by including the delayed component that TPB exhibits when wavelength-shifting the VUV argon scintillation light, and estimate the fraction of surface alpha background events that may fall into this adjusted acceptance region. Due to the lack of argon NR data with millisecond-long digitization, this is our best effort to incorporate the effect of the delayed component in surface-coated TPB and bulk-dissolved TPB~\cite{ref:Asaadi}.

Because the TPB response to 128\,nm photons has a relatively small delayed component, the choice of $W$ only makes a small impact on the pulse integral and the $f_{90}$ parameter for dark matter NR signals. 
The magnitude of this effect is obtained by fabricating pure NR pulses composed of two exponentials, corresponding to the singlet and triplet argon lifetimes, and convolving them with the fit ($I_{fit}(t)$) to the AW of Measurement C (see Appendix~\ref{sec:model}):
$$I_{NR}(t)=I_{fit}(t)\ast(N_Se^{-t/7\mathrm{ns}}+N_Te^{-t/1.6\us})$$
where the singlet/triplet ratio ($N_S/N_T$) can be adjusted to produce pulses of arbitrary $f_{90}$ and thereby produce a correction for $f_{90}$ values calculated with a different $W$. The lifetimes of the singlet and triplet states were obtained from~\cite{ref:ArLifetimes}. While other published measurements of the triplet lifetime vary, the uncertainty of this value was determined to have a negligible impact on this correction.

\begin{figure}[tp]
\includegraphics[width=0.5\textwidth]{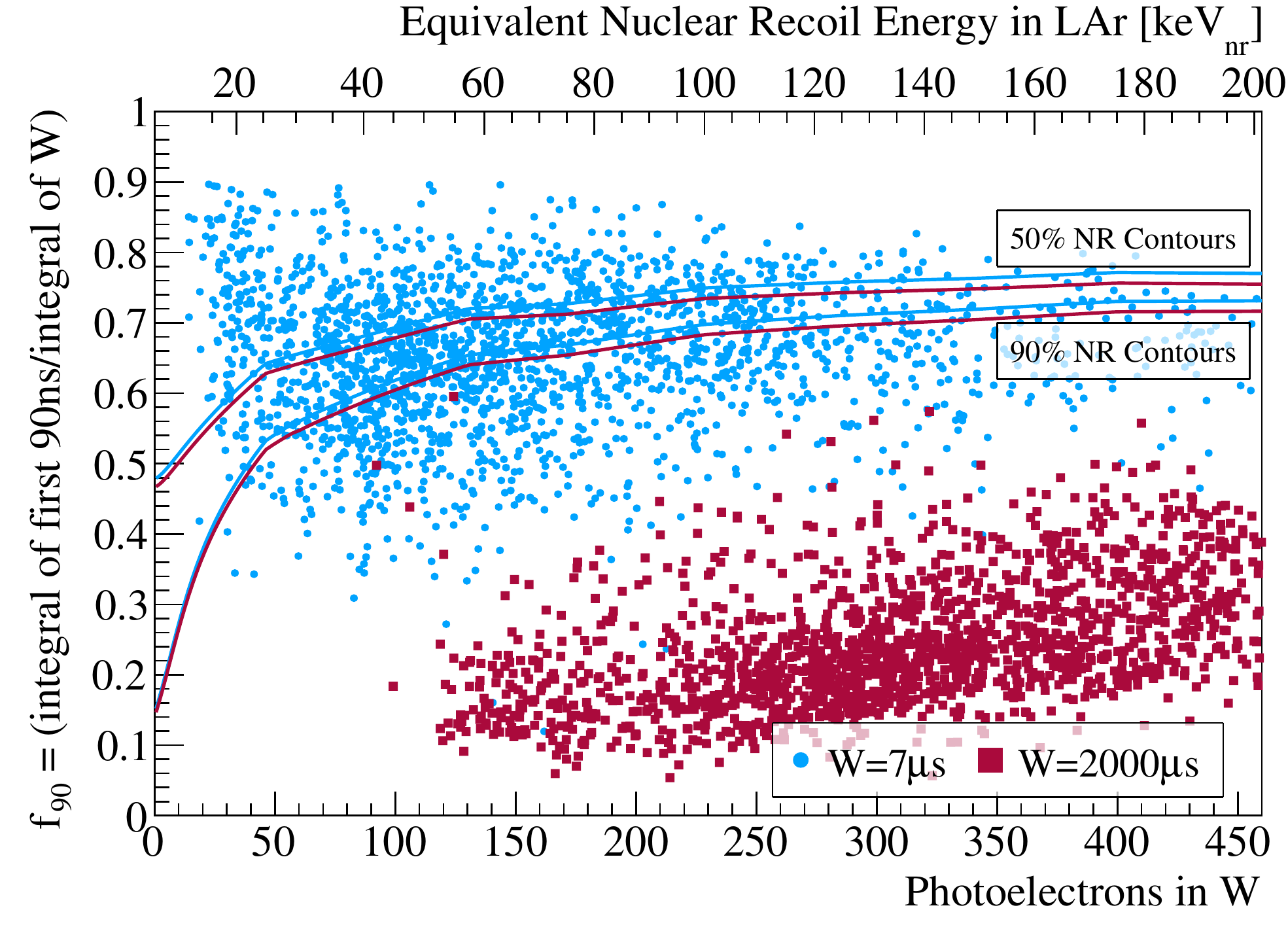}
\caption{The effect of the integration window size $W$ on the pulse shape parameter $f_{90}$ of surface alpha decays. By using a larger window, more of the photoelectrons from the delayed component are included, and the events are moved away from the nuclear recoil contours obtained from~\cite{ref:DS70day}. The nuclear recoil energy scale is also obtained from~\cite{ref:DS70day}.}
\label{fig:longtail_compare}
\end{figure}

After recalculating the $50\%$ and $90\%$ NR contours from \cite{ref:DS70day} for a larger $W$, we observed a slight shift downward, as expected. These curves can be seen in Fig.~\ref{fig:longtail_compare}.

On the other hand, due to the magnitude of the delayed component of the TPB scintillation under alpha excitation, the region of $f_{90}$-NPE space for surface background events is heavily dependent on $W$. A larger choice of $W$ will result in a larger overall pulse integral and thus a smaller $f_{90}$, away from the dark matter NR region. 
Fig.~\ref{fig:longtail_compare} shows an example of this effect, where the same surface background events are plotted with two different $W$ values: $W=7$\us\ and $W=2000$\us. 
In order to properly compare results across the two experiments, the pulse integrals from our experiment were scaled up by a factor of $7.90/5.87$, based on the the difference in the measured light yield and argon purity (triplet lifetime). 
The systematic errors introduced by these corrections have been propagated into the final results. 

With the 7\us\ integration window (as used in DarkSide-50), \percentsevenus\% of $\sim$2k surface alpha decays produced a signal above the 90\% contour. With the 2000\us\ integration window, no events were observed above the contour, resulting in a suppression factor of over $10^\rejecttwomspower$. We emphasize that this suppression comes at no cost to a detector's fiducial volume, so it can significantly enhance the sensitivity of a LAr experiment that had been using position cuts to remove surface backgrounds. In Section~\ref{sec:other}, we will discuss possible applications of this technique beyond just LAr experiments.

\begin{figure}[tp!]
\includegraphics[width=0.5\textwidth]{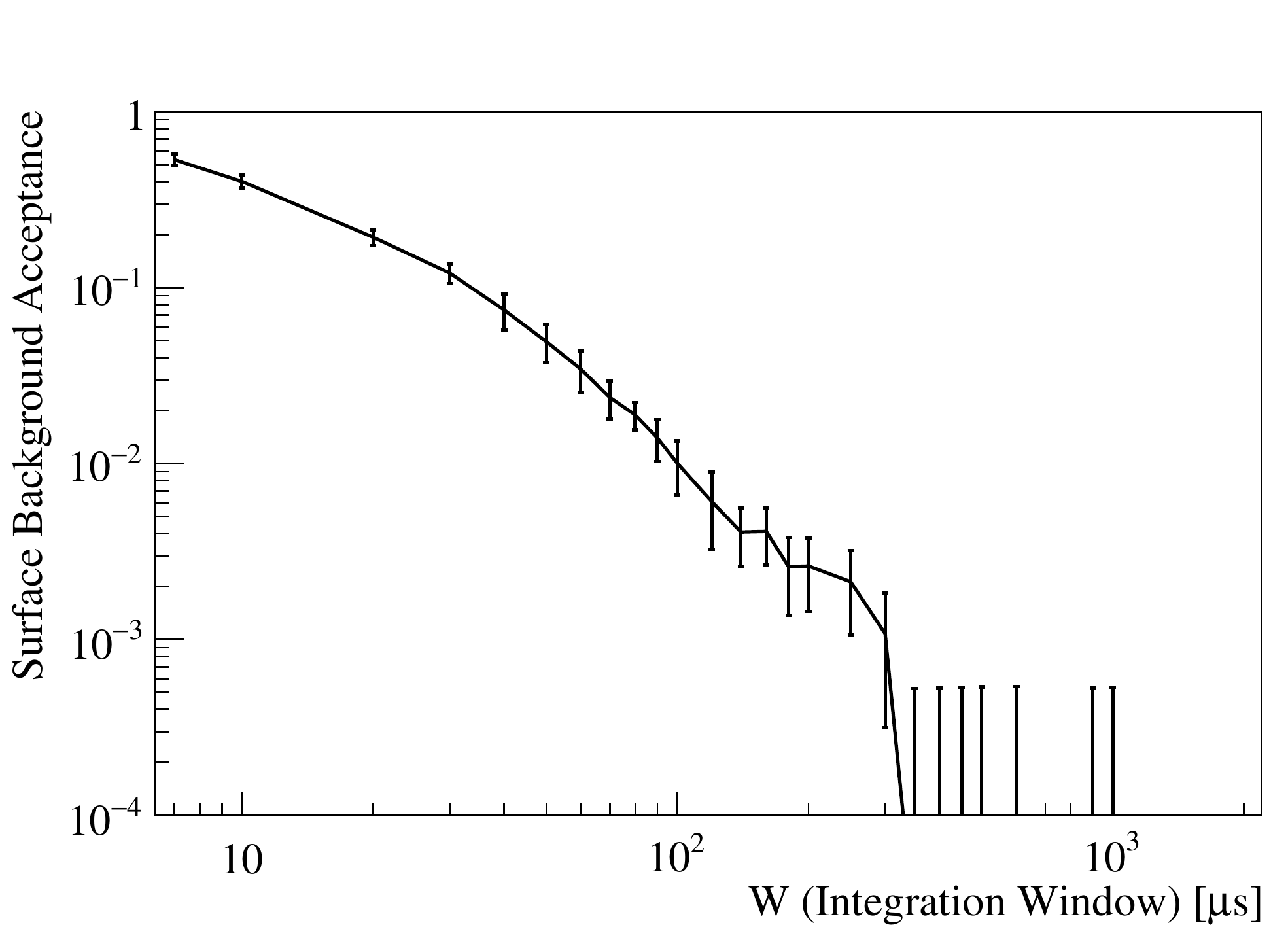}
\caption{The fraction of surface alpha decays that would fall above a 90\% nuclear recoil acceptance contour as a function of integration window size.}
\label{fig:longtail_acc}
\end{figure}

For experiments with high trigger rates that are concerned about data volume and pileup events, the use of milliseconds-long acquisition windows may be impractical. Therefore, we extended the estimation of the surface background acceptance to different choices of $W$, which can be found in Fig.~\ref{fig:longtail_acc}. The error bars include statistical errors as well as the systematic errors introduced by the comparison to DarkSide-50. 

Since the radon progeny in this study were deposited on top of the TPB, the background suppression above applies to surface decay mode I(b) from Fig.~\ref{fig:decay_kin}. However, since it features a NR component in the LAr, it is expected to be the most difficult mode for a LAr experiment to suppress. Back in Section~\ref{sec:decaymodes}, we found that each other mode that poses a risk of background would have a similar or larger fraction of its scintillation come from the TPB, and would therefore have an even more prominent delayed component. So this measurement can be considered to be a conservative estimate of the achievable suppression of all surface backgrounds modes.

We also determined that our use of \pofour, which produces higher energy alphas (7.8 MeV) and nuclear recoils (146 keV) than \poten, results in a conservative estimate of the power to suppress \poten\ background. 
This is because the most concerning surface background for a WIMP experiment occurs at low energies when the alpha is ejected at an angle directly into the wall and only deposits a small fraction ($\sim200$\,keV) of its energy in the TPB. 
In such cases, due to the larger $\abs{\dd E/\dd x}$ in TPB of \poten\ alphas, they actually deposit more energy in the thin layer of TPB than \pofour\ alphas, as confirmed by SRIM simulations~\cite{ref:SRIM}. As a result, \poten\ is expected to produce more TPB scintillation (assuming TPB obeys Birks' law~\cite{ref:Birks1951}) and less LAr scintillation (due to a lower energy recoil) than \pofour, making it even easier to suppress.

Finally, we point out that while using $f_{90}$ as a pulse shape parameter was necessary to make the comparison to DarkSide-50, experiments may find that other choices offer even greater levels of suppression---such as the fraction of light that occurs more than 10\,\us\ after the prompt.

\section{OTHER APPLICATIONS\label{sec:other}}

Since the delayed component of TPB scintillation was found to be present across a wide temperature range, its usefulness as a surface background veto can be extended beyond just LAr experiments. For example, a scintillating crystal could be coated in TPB to act as a veto for alphas ejected from the surrounding detector components.

For experiments that are not suited for TPB, alternative scintillating coatings may be used to suppress surface backgrounds.
While the detailed results are beyond the scope of this work, we discovered two other compounds that had delayed scintillation components similar to TPB. The first compound was p-Terphenyl, another organic WLS, which suggests that a milliseconds-long scintillation component may be a property of aromatic scintillators in general. The second compound was MgF$_2$, an inorganic material. MgF$_2$ may be more suitable for liquid xenon experiments, which tend to dissolve organic compounds like TPB and p-Terphenyl~\cite{ref:TPBxenon,ref:PTPxenon}. MgF$_2$ is also transparent to liquid xenon scintillation light. 
Further investigation into these compounds is ongoing. 

\section{CONCLUSIONS}

We have demonstrated the existence of a delayed component of TPB scintillation that has a lifetime on the order of milliseconds, and that the magnitude of this component depends strongly on the type of incident radiation. Scintillation pulses caused by alpha particles contain the majority of their light in the delayed component. This allows for pulse-shape discrimination to be used to reject surface backgrounds. When compared to the nuclear recoil acceptance contours measured by DarkSide-50, a surface background suppression factor of greater than $10^\rejecttwomspower$ can be obtained when using milliseconds-long integration windows.

\section{ACKNOWLEDGMENTS}
We thank Ben Loer for developing the data acquisition software that was used in this experiment. We are grateful to Peter Meyers for his insightful suggestions on the analysis. This work was supported by the NSF grants PHY0704220 and PHY0957083. J.Xu is an employee of the Lawrence Livermore National Laboratory (LLNL). LLNL is operated by Lawrence Livermore National Security, LLC, for the U.S. Department of Energy, National Nuclear Security Administration under Contract DE-AC52-07NA27344.

\appendix

\section{TPB SCINTILLATION MODEL\label{sec:model}}

The long tails of the AWs do not follow an exponential decay process. Rather, the shape resembles a power-law decay with an exponential cutoff. In\ \cite{ref:LongTailTheory} (English translation \cite{ref:thesis_chris_stanford_translation}), R. Voltz and G. Laustriat describe this behavior using a model of the scintillation process of organic compounds which considers both the prompt and delayed response of the scintillator. The following is a brief summary of their model.

When a source of ionizing radiation is absorbed by the scintillator, the molecules are excited into singlet and triplet states. The singlet states decay quickly while the triplet states have a longer decay lifetime and produce a ``delayed" signal. This process is not represented simply by an extra exponential term. In regions of the scintillator that have received high levels of ionization, such as the track produced by an alpha, intra-molecular interactions become significant; triplet states can be destroyed through triplet-triplet annihilation with timescales related to the diffusion of these dense pockets of ionized species.

The full equation describing the instantaneous intensity of the scintillation versus time takes the form: 
\begin{equation}
I(t)=I_p(t)+I_d(t)
\label{eq:I}
\end{equation}
where $I_p(t)$ and $I_d(t)$ represent the instantaneous intensities of the prompt and delayed components, respectively. $I_p(t)$ is a result of the decay of the short-lived singlet states and is represented by an exponential:

\begin{equation}
I_p(t)=\frac{N_p}{\tau_S}\exp(-t/\tau_S)
\end{equation}
where $N_p$ is the total integrated intensity of $I_p(t)$ and $\tau_S$ is the lifetime of the singlet states.

The instantaneous intensity of the delayed component, $I_d(t)$, is derived by considering the diffusion of densely packed triplet states and their interactions with each other. The full form of $I_d(t)$ is cumbersome, so we use a simplified expression that is valid for $t\gg\tau_S$, with the justification that the small inaccuracies of $I_d(t)$ expected at short timescales can be ignored due to the fact that $I_p(t)$ is dominant in the $t\sim\tau_S$ regime by orders of magnitude. The simplified form is:

\begin{equation}
I_d(t)=\frac{N_d}{\tau_d} F(t)
\end{equation}

Where $N_d$ is the total integrated intensity of $I_d(t)$ and $\tau_d=\int_0^{\infty} F(t) \dd t$ is a normalization constant. $F(t)$ is defined as:

\begin{equation}
  F(t)=\frac{e^{-2t/\tau_T}}{\left \{1+A\bigg[\Ei(-\frac{t+t_a}{\tau_T})-\Ei(-\frac{t_a}{\tau_T})\bigg] \right \}^2(1+t/t_a)}
  \label{eq:F}
\end{equation}
Where: 
\begin{itemize}
\item $\tau_T$ is the lifetime of the triplet state,
\item $t_a$ is a time constant defined as $t_a=r_0^2/4D_T$, where $r_0$ is the characteristic width of the ionization track and $D_T$ is the diffusion coefficient of the triplet states,
\item $A=\frac{t_a}{2t_b}\exp(\frac{t_a}{\tau_T})$, where $t_b=[\chi_{tt}C_T(0)]^{-1}$, with $\chi_{tt}$ being the rate constant for bimolecular quenching of triplet states and $C_T(0)$ is the initial concentration of triplet states,
\item  $\Ei(-x)=-\int_x^\infty\frac{e^{-\alpha}}{\alpha}\dd\alpha$ is the exponential integral.
\end{itemize}

Visually, $F(t)$ looks flat up until $t_a$, then it transitions to a power-law decay, which appears as a straight line on a log-log plot, with the slope affected by $A$. After reaching $t\approx\tau_T$, the exponential term begins to dominate, cutting off the power-law. 

The formula $I(t)$ given by Voltz and Laustriat needed a few empirical additions before it could be fit to the AWs. 


The first additional term was due to the smearing caused by the PMT response. The response of the PMT was obtained using a HORIBA NanoLED pulsed picosecond laser of similar wavelength to the TPB emission spectrum. This response $R(t)$ was then convolved with the model.

The second was due to a flat background rate of single-photoelectron PMT ``dark noise". This introduced a constant term $C$. When fit, this term was constrained to be within 1$\sigma$ from a constant fit to the 50\us\ pre-prompt region of the AW. To avoid biasing this value, the trigger threshold for the data acquisition was set at a level that would not trigger on single photoelectrons.

The full fit function is thus:

\begin{equation}
\begin{aligned}
I_{fit}(t)=R(t)*\Big(&I_p(t;N_p,\tau_S) \\
+&I_d(t;N_d,A,t_a,\tau_T)\Big)+C 
\end{aligned}
\label{eq:fullfunc}
\end{equation}

Since $C$ does not represent any component of the TPB scintillation, this term was subtracted after the fit and a re-scaling was performed on each of the integrated intensity terms such that $N_x'=N_x/(N_p+N_d)$ for $x=p,d$. Since each waveform was normalized to a value of 1, doing this allowed $N_p'$ and $N_d'$ to be interpreted as the faction of light of the TPB scintillation that was due to the prompt and delayed components, respectively.

In the case of the 128\,nm photon measurement (Fig.~\ref{fig:schematics}C), the prompt exponential $I_p(t)$ actually represents the convolution of the argon prompt scintillation with the TPB prompt response. The argon prompt lifetime is not known to high enough precision to warrant deconvolving these responses~\cite{ref:ArLifetimes}.

In the case of the surface background measurement (Fig.~\ref{fig:schematics}D), the AW could not be fit by $I_{fit}(t)$. This is because at $t<10\us$ there are additional components from the argon scintillation of the recoiling nucleus, which are not included in the model.

The fitted parameter values for each AW can be found in Table~\ref{tab:parameters}.
There are several observations to be made regarding the fits.


\begin{table*}
\centering
\begin{tabular}{@{}l|l|l|l|l|l|l@{}}
\toprule
Description & \multicolumn{1}{c}{$N_p'$} & \multicolumn{1}{c}{$\tau_S$(ns)} & \multicolumn{1}{c}{$N_d'$} & \multicolumn{1}{c}{$A$}      & \multicolumn{1}{c}{$t_a$(\us)}  & \multicolumn{1}{c}{$\tau_T$(\us)} \\ \midrule

Betas @ 295K & \betaWarmNp(\betaWarmNpErr) & \betaWarmTaus(\betaWarmTausErr) & \betaWarmNd(\betaWarmNdErr) & \betaWarmA(\betaWarmAErr) & \betaWarmTa(\betaWarmTaErr) & \betaWarmTaut(\betaWarmTautErr)$\times10^2$ \\

Alphas @ 295K & \alphaWarmNp(\alphaWarmNpErr) & \alphaWarmTaus(\alphaWarmTausErr) & \alphaWarmNd(\alphaWarmNdErr) & \alphaWarmA & \alphaWarmTa(\alphaWarmTaErr) & \alphaWarmTaut(\alphaWarmTautErr)$\times10^2$ \\

Betas @ 77K & \betaColdNp(\betaColdNpErr) & \betaColdTaus(\betaColdTausErr) & \betaColdNd(\betaColdNdErr) & \betaColdA(\betaColdAErr) & \betaColdTa(\betaColdTaErr) & \betaColdTaut(\betaColdTautErr)$\times10^3$ \\

Alphas @ 77K & \alphaColdNp(\alphaColdNpErr) & \alphaColdTaus(\alphaColdTausErr) & \alphaColdNd(\alphaColdNdErr) & \alphaColdA & \alphaColdTa(\alphaColdTaErr) & \alphaColdTaut(\alphaColdTautErr)$\times10^3$ \\

128nm photons & \argonColdNp(\argonColdNpErr) & \argonColdTaus(\argonColdTausErr) & \argonColdNd(\argonColdNdErr) & \argonColdA(\argonColdAErr) & \argonColdTa(\argonColdTaErr) & \argonColdTaut(\argonColdTautErr)$\times10^3$ \\
\bottomrule
\end{tabular}
\caption{The parameters for $I_{fit}(t)$ when fit to the average waveforms for each measurement. The total errors, which are dominated by systematics, are shown in parentheses, with $n$ digits in parentheses indicating the error on the last $n$ digits.} 
\label{tab:parameters}
\end{table*}

The fits for alphas at room temperature and \lntwo\ temperature have $N_d'\sim0.8$, indicating that in these cases the delayed component is in fact the dominant component. 

For betas, the delayed component was significantly smaller, but still non-negligible with $N_d'\sim 0.3$. Therefore, while the focus of this paper was on surface alpha decays, it appears that TPB could be used to suppress surface beta decays as well. Beta particles have a higher penetrating power, and often deposit only a small fraction of their total energy as they pass through the TPB ($\sim0.1\%$ for 500\,keV betas in 2\um-thick TPB~\cite{ref:ESTAR}), so they produce a smaller absolute pulse size compared to alphas. In practice, a relatively thick layer of TPB might be required to suppress surface betas.  

This difference in delayed component magnitude ($N_d'$) between different particle species is well explained by the Voltz-Laustriat model. When Eq.~\ref{eq:F}, which describes the delayed component, is simplified by assuming $A\ll 1$ (true for many of our fits), the integral can be obtained analytically:

\begin{equation}
  \int_0^\infty F(t)\approx -t_a e^{2 t_a / \tau_T}\Ei(-2 t_a/\tau_T)
\end{equation}
which increases monotonically with $t_a$. Recalling that $t_a$ is proportional to the square of the ionization track width, we would expect the wider ionization tracks left by alphas to result in a comparatively greater delayed component magnitude.

In contrast, the variation of $\tau_T$ between the alpha and beta measurements is not predicted by the model. While interesting, we draw no conclusions because the alpha and beta measurements were performed at different times with different TPB samples, and the resulting systematics that might affect $\tau_T$, such as the sample thickness, age, and purity, were not considered in this study. While we limited the exposure of our samples to ambient air and UV, an experiment with foreknowledge of the delayed component that is dedicated to completely avoiding these sources of TPB degradation may be able to make a better comparison between samples, and may see an even larger $\tau_T$ and an even greater level of surface background rejection than we report here.

For both alphas and betas, there was an observed temperature dependence. While the fraction of light in the delayed component remained nearly constant, the observed triplet lifetime $\tau_T$ increased by an order of magnitude when going from warm to cold. This is consistent with observations of temperature dependence of phosphorescence reported in~\cite{ref:phosphorTempDep}.

Lastly, not shown in this paper, was an observed energy dependence of the AW, most notably for alphas. Alphas that pass normally through the TPB deposit less energy than alphas that pass through at a steep angle. If an AW is produced for different slices of the energy spectrum, slight differences can be seen in the delayed component. The AWs we present here are weighted sums over all energies, which correspond to probabilistic sums over all solid angles. This represents the average signal expected from a decay, which can eject its products in any direction. However, it does mean that the model should not be expected to fit the data perfectly, since it assumes a constant stopping power (and resulting initial triplet state concentration $C_T(0)$) over the length of the particle's track. 

\section{TPB ALPHA LIGHT YIELD\label{sec:TPBLY}}

Future LAr experiments may wish to simulate alpha decays on surfaces with an arbitrary thickness of TPB. To do this, an absolute scale for the scintillation light yield of alphas in TPB must be established, which we provide here using the alpha decays from Measurement A. The other component of the surface decay, namely the recoiling nucleus in LAr, is characterized in~\cite{ref:PbRecoil}. These measurements, taken together, can enable full simulations of alpha decays in LAr on TPB surfaces of any thickness. 

\begin{figure}[t!]
\includegraphics[width=0.5\textwidth]{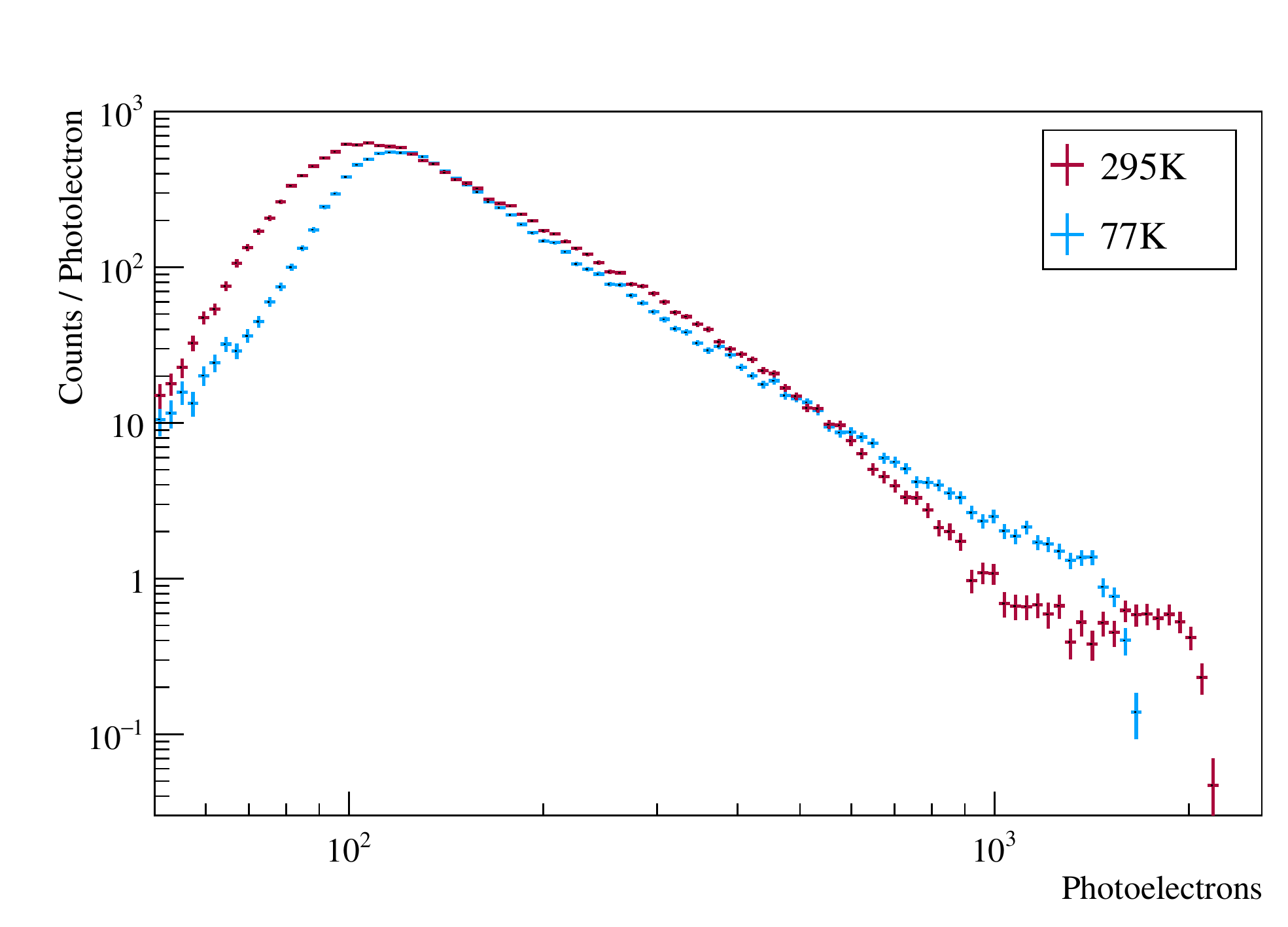}
\caption{The energy spectrum for alpha scintillation in TPB (Measurement A). The spectrum has been corrected to account for integration window size. The alpha source was placed near the TPB layer, so alphas can be observed passing through at small angles (relative to normal), resulting in the peak at $\sim$100 photoelectrons, or a large angles, resulting in the peak at $\sim$2000 photoelectrons. The lack of the higher energy peak in the 77\,K data is not understood.}
\label{fig:po210_tpb_spectrum}
\end{figure}

The energy spectrum for Measurement A can be found in Fig.~\ref{fig:po210_tpb_spectrum}. An integration window of 2~ms was used to determine the energy, then the fits to the pulse shapes from Appendix~\ref{sec:model} were used to correct the energy to what would be expected with an integration window of infinite length.

The events in the 100~PE peak correspond to alphas that left the \poten\ needle source at a trajectory normal to the TPB surface, depositing a small fraction of their energy as they passed through. The highest energy events in the spectrum correspond to alphas that entered the TPB surface at a steep angle and deposited their full energy of 5.3~MeV. A peak forms here because alphas at even greater angles of incidence cannot deposit any more energy. The lack of such a peak in the 77~K data is not understood. Using the peak at 2000\,PE gives a light yield of 0.38$\pm$0.02~PE/keV at 295~K. If we assume that the ratio of light yields is constant over the measured energy range, the ratio of the low energy peaks (($116\pm 2$~PE)/($108\pm 2$~PE)) then gives a light yield of 0.41$\pm$0.03~PE/keV at 77~K.

Note that a simplification has been made here that the alpha light yield is independent of stopping power, when it is likely that the light yield changes along the ionization track as a function of $\dd E/\dd x$. Resolving these non-linearities would require additional measurements.

To enable the translation of these values to other experiments, we also provide our LAr light yield of 5.87$\pm$0.37~PE/keV for a 60~keV gamma source as a measure of our light collection efficiency. 

Our finding that the total scintillation yield increases slightly from warm to cold temperatures disagrees with a previous experiment that measured a decreasing light yield with colder temperatures~\cite{ref:TPBTemp}. This discrepancy can be resolved with our measurement of the delayed component lifetime, which showed an increase from \alphaWarmTautus\us\ at 295\,K to \alphaColdTautus\us\ at 77\,K. Using a shorter acquisition window, such as the 200\us\ window used in the cited experiment, may result in the appearance of decreasing light yield at colder temperatures since more of the light would fall past the end of the acquisition window. Our measurement, which corrects for the choice of acquisition window, is expected to be more accurate.

\bibliographystyle{ds}
\bibliography{mybib}

\end{document}